\documentclass[sigconf]{acmart}

\AtBeginDocument{%
  }

\copyrightyear{2025}
\acmYear{2025}
\setcopyright{rightsretained}
\acmConference[CHI EA '25]{Extended Abstracts of the CHI Conference on
Human Factors in Computing Systems}{April 26-May 1, 2025}{Yokohama, Japan}
\acmBooktitle{Extended Abstracts of the CHI Conference on Human Factors in
Computing Systems (CHI EA '25), April 26-May 1, 2025, Yokohama,
Japan}
\acmDOI{10.1145/3706599.3719987}
\acmISBN{9 7 9 - 8 - 4 0 0 7 - 1 3 9 5 - 8 / 2 0 2 5 / 0 4 }

\usepackage{makecell}

\begin{document}

\title{Under Pressure: Contextualizing Workplace Stress Towards User-Centered Interventions}


\author{Antonin Brun}
\affiliation{%
  \institution{University of Southern California}
  \city{Los Angeles}
  \state{California}
  \country{USA}}
\email{abrun@usc.edu}

\author{Gale Lucas}
\affiliation{%
  \institution{University of Southern California}
  \city{Los Angeles}
  \state{California}
  \country{USA}}
\email{lucas@ict.usc.edu}

\author{Bur\c{c}in Becerik-Gerber}
\affiliation{%
  \institution{University of Southern California}
  \city{Los Angeles}
  \state{California}
  \country{USA}}
\email{becerik@usc.edu}

\renewcommand{\shortauthors}{Brun et al.}

\begin{abstract}
    Stress is a pervasive challenge that significantly impacts worker health and well-being. Workplace stress is driven by various factors, ranging from organizational changes to poor workplace design. Although individual stress management strategies have been shown to be effective, current interventions often overlook personal and contextual factors shaping stress experiences. In this study, we conducted semi-structured interviews with eight office workers to gain a deeper understanding of their personal experiences with workplace stress. Our analysis reveals key stress triggers, coping mechanisms, and reflections on past stressful events. We highlight the multifaceted and individualized nature of workplace stress, emphasizing the importance of intervention timing, modality, and recognizing that stress is not solely a negative experience but can also have positive effects. Our findings provide actionable insights for the design of user-centered stress management solutions more attuned to the needs of office workers.
\end{abstract}




\maketitle

\section{Introduction}
Coined as the “modern day epidemic” \cite{kalia_assessing_2002}, work-related stress is a prevalent issue that extends beyond the workplace. In the US alone, 83\% of workers experience work-related stress, and 120,000 die each year from complications related to workplace stress \cite{osha_workplace_nodate}. Work-related stress not only affects worker well-being and productivity but also frequently extends into their home life \cite{wepfer_work-life_2018, the_american_institute_of_stress_workplace_2025}. Stress also has significant ramifications on mental health, increasing the risk for anxiety and depression, and contributing to disrupted sleep and fatigue \cite{guilliams_chronic_2010}. Workplace stress can be driven by a variety a factors, from organizational changes and work commitments \cite{mark_email_2016, mark_stress_2014, mark_how_2017, mcduff_longitudinal_2021}, to poor indoor environmental quality (IEQ) conditions \cite{Choi_2015, sander_open-plan_2021, Thach_2020, Toftum_2012}. Consequently, there has been a growing interest in providing workers with effective stress management strategies.

To effectively address workplace stress, individual stress management strategies (e.g., cognitive-behavior interventions, breaks, physical activity, etc.) are an effective approach \cite{cooper_intervention_1997, richardson_effects_2008}. With the advancement of multimodal sensing \cite{sander_open-plan_2021, booth_toward_2022} and machine learning \cite{chodan_seba_2020}, there has been a growing interest within the Human-Computer Interaction (HCI) community to develop and design workplace stress management interventions. While some interventions \cite{haliburton_walking_2023} have positive effects on workplace stress by extension, our focus, however, is on interventions directly tailored to the physical workplace context. Such interventions leverage multimodal sensing (physiological/behavioral data from wearables, keyboards, and mice) \cite{androutsou_smart_2021, androutsou_multisensor_2023, sun_moustress_2014, pepa_stress_2021, booth_toward_2022}, Just-in-Time Adaptive Interventions (JITAIs) \cite{suh_toward_2024, howe_design_2022, clarke_mstress_2017, lustrek_designing_2023}, to detect and help workers reduce their stress. The integration of these interventions with workplace stress theories \cite{kowatsch_towards_2015, koldijk_deriving_2016, nixon_efficacy_2022} allow us to shape the interventions to the workers' needs. However, the effectiveness of these interventions can be influenced by personal factors such as demographics and individual stress experiences, aspects that are sometimes overlooked in current research. While prior HCI studies have explored the role of personal factors in workplace well-being (e.g., technological implications of cultural differences in perceptions of workplace norms \cite{akahori_impact_2024}, and speculative design addressing the tensions between home and office work with remote/hybrid work arrangements \cite{cho_reinforcing_2024}), these factors are not always directly considered when designing stress management interventions.


In this study, we aim to further contextualize stress in the modern workplace space. Through a set of semi-structured interviews with eight office workers, we gain rich insights into their personal experiences with stress. We explore stress triggers, remediation techniques and coping mechanisms, as well as reflections on past stressful events. We highlight that worker stress is inherently multifaceted and worker-specific. We also discuss the value of intangible interfaces integrated with workplace design, the importance of intervention timing and modalities, and the dual-nature of stress which can sometimes be beneficial to workers. This work provides insights into the future of workplace stress management and how we can design more effective and user-centered interventions.

\section{Background}
In this section, we first explore workplace stressors, and we then review interventions for managing workplace stress. 

\subsection{Stressors in the workplace}
Workplace stress remains a critical concern within organizational psychology and occupational health research. Specifically, office workers face unique stressors tied to the psychosocial and physical contexts of their roles. Stress in office environments arises from a combination of organizational and interpersonal factors (e.g., workload, role ambiguity, management issues), individual-level moderators (e.g., coping mechanisms), and environmental conditions (e.g., workspace design, noise, lighting) \cite{bolliger_sources_2022, june_effect_2013, herbig_does_2016, norouzianpour_architectural_2020, colligan_workplace_2006, lindberg_effects_2018, Thach_2020, Toftum_2012, sander_open-plan_2021, Choi_2015}. Psychosocial stressors are one of the main contributors to workplace stress. These include high workloads and time pressure \cite{bolliger_sources_2022, colligan_workplace_2006}, job insecurity and effort-reward imbalances \cite{bhui_perceptions_2016}, organizational culture shifts and inadequate managerial or social support \cite{bhui_perceptions_2016, bolliger_sources_2022}. Alongside psychosocial dimensions, environmental stressors also have a significant impact on worker stress. Office design \cite{lindberg_effects_2018}, noise \cite{Toftum_2012, sander_open-plan_2021, Choi_2015}, lighting \cite{Thach_2020}, and overcrowding \cite{herbig_does_2016} have been found to exacerbate stress. Furthermore, rapidly growing technological advancements and perpetual connectivity – often referred to as technostress – have also been associated with increasing workplace stress \cite{vold_new_1987, bolliger_sources_2022}. These insights highlight the complex nature of workplace stress, showing that it results from a combination of factors rather than a single cause.

\subsection{Workplace stress interventions}

Workplace stress can be managed through a variety of strategies. Individual management strategies are regarded as most effective in the modern workplace \cite{cooper_intervention_1997, richardson_effects_2008}. This recognition has spurred extensive research in personal informatics and HCI to develop tools aimed at reducing worker stress \cite{hickey_smart_2021}. Most of these solutions rely on web-based and desktop-embedded \cite{howe_design_2022, sano_designing_2017, tong_just_2023} or touchscreen-based personal interfaces \cite{ahire_workfit_2024}. While the overarching goal of managing stress is consistent, the methods vary. Some approaches focus on helping users reflect on their stress and prepare for high-pressure situations \cite{lee_toward_2020}, while others emphasize mindfulness exercises \cite{tong_just_2023} or suggest tailored interventions via a digital personal desktop assistant \cite{howe_design_2022} our through wearable technologies \cite{ahire_workfit_2024}. 


The current landscape of workplace stress interventions reveals several limitations. First, these strategies often rely on active user participation and visual interactions. These interventions might not be suited for all workers or scenarios –  they can ultimately increase extraneous cognitive load \cite{sweller_chapter_2011} and aggravate already stressful situations. Second, the balance eustress-distress \cite{le_fevre_eustress_2003} is often disregarded when designing these interventions, thus overlooking the potential to promote healthier work habits. Third, most of the research focusing on active interventions aims at reducing existing stress (e.g., prompting to take breaks, therapy-based approaches) often neglects the diverse range of stressors and experiences in the workplace that could inform more effective preventive measures. We emphasize the need for qualitative, context-driven studies to better understand workplace stress and enhance the design of stress management interventions.

\section{Methods}

To explore how office workers experience stress in the workplace, we conducted elicitation interviews with eight participants. These individuals were previously involved in a longitudinal study focused on detecting and mapping workplace stress. Our interviews served as a follow-up to the original study and took place after participants had completed their participation in the longitudinal research. During the interviews, participants were asked to recall specific stressful events they experienced within the past few months. All procedures were reviewed and approved by our University’s Institutional Review Board (IRB).

This study is the continuation of a broader research initiative where multimodal data was gathered from 15 participants over a 4-month period. We collected the participants’ physiological data using a FitBit watch, their levels of social interaction using an audio recording device, as well as continuous IEQ measurements of their office spaces. Throughout the data collection, we also captured the participants’ perceived physical and mental health symptoms, stress, mood, and productivity levels, using semi-random momentary ecological assessments (EMAs) (i.e., short questionnaires).

We defined "office workers" as individuals who primarily perform their work at an assigned desk (excluding hot-desking roles). Participants were recruited based on their psychosocial job characteristics using the Job Content Questionnaire (JCQ) \cite{karasek_job_1998} and eligibility criteria, which included being over 18 years old at the time of the study. All participants were affiliated with the University of Southern California (USC). Out of 15 individuals who completed the longitudinal data collection, two were not affiliated with USC anymore, leaving us with 13 eligible participants for interviews. A total of eight participants responded to our solicitations. Participant demographics and psychosocial job characteristics are detailed in Table \ref{tab:demographics}.

We used a semi-structured interview format designed to capture participants' experiences of workplace stress. Specifically, our approach draws on the “story interview method”, a story-based design technique grounded in HCI theory. Story interviews are particularly valuable for capturing rich, detailed information about user experiences, enabling researchers to derive actionable insights for design \cite{mackay_doit_2023}. Initially, participants were asked to recall specific stressful events that they had encountered in the past few months. The participants were then asked to walk us through a stressful situation that they had recently experienced. To better understand their experiences, we asked three focused questions: (1) What led them to feel stressed? (2) Did they take any actions to manage or reduce their stress? (3) Did they share their feelings or experiences with others? Participants were encouraged to elaborate further as they walked us through their experiences. Interview questionnaire is detailed in Appendix \ref{itw_guide}. The interview data were analyzed using thematic analysis \cite{braun_using_2006} to identify patterns and themes in participants’ descriptions of workplace stress. 

\section{Findings}
In this section, we report on our insights from the interviews and contextualize stress in the workplace. 

\subsection{Main stressors}

We highlight the main causes of stress in the workplace reported by our participants. 

\subsubsection{Organizational constraints}

Most participants experienced stress due to competing demands within their organizations. Organizational pressures often came from multiple stakeholders requiring attention, with limited resources or time to meet all demands. Two participants specifically discussed hierarchical dynamics as their main stressor. These dynamics can take the form of jarring disruptions, as P8 reported that her supervisor is \textit{“always constantly on the go and busy [...] and likes to be able to get answers immediately and quickly,”} or unreasonable work demands as P3 highlighted, \textit{“[They’ve] basically given me 2 days to create 10 years of data out of nothing.”} Organizational stress, however, does not always come from the higher-ups. For example, when describing a recent stressful event, P7 noted a lack of training resources as the main source of stress, stating that her undertrained colleague \textit{“walked into the room and just pressed all the buttons, ruining the microphones and the Zoom setup,”} quickly escalating the participant’s stress levels.

\subsubsection{Workplace design} \label{workplace_design}

Three participants specifically cited the physical workplace as both a source of stress and a source of relief. Lack of access to natural light were highlighted as stressors for two participants. For example, while giving an online lecture, P2 reported, \textit{“if I'd had a window, I would have been able to look out and like, breathe. And I was just feeling so tied to the virtual space of the workstation because [the students] are online,”} which ultimately worsened her already stressful experience \textit{“I think it just sort of compounded it.”} In contrast, other participants took advantage of their workplace design to increase their comfort and manage their stress. P4 stated, \textit{“[I] tend to keep my [partition] windows closed in the office so I'm very isolated”} as a means to minimize disruptions and unnecessary stress, which highlights the importance of individual preferences in managing stress.

\subsubsection{External factors}

The impact of external factors (i.e., other than work commitments and workplace design) were mixed. Two participants specifically talked about external factors affecting their stress. When discussing her personal experience, P2 highlighted that personal issues spilled over into the workplace, affecting her stress, \textit{“Yeah, and just like life things right now. At the time my husband was still looking for a job. He's been unemployed. And so you know how things in your life can be like.”} On the other hand, another participant reported that his personal life does not intrude into his work life. P1 stated that he is able to manage his time and commitments both for work- and non-work-related issues, \textit{“I feel like I know I'm going to get somewhat of a small break. So other commitments outside of work, family, or whatever else I think I was on a good level of managing it.”} 

\subsection{Stress remediation strategies}

We discuss the various techniques used by office workers to manage their stress.

\subsubsection{Physical and mental breaks}

The vast majority of participants (six out of eight) reported taking breaks to manage their stress. These pauses took different forms. Some participants opted for breaks involving physical activity to dissociate themselves from the stressful task at hand (usually associated with their workplace). For example, P7 highlighted that her office location made it convenient for her to go outside for a break: \textit{“our office is across the street from the marina, and I just went outside for a walk. It's a 20-minute walk around Marina del Rey.”} On the other hand, P2 discussed purposefully walking outside despite lacking access to a dedicated walking area. She stated, \textit{“I walked out to my car and then walked back,”} echoed by how she usually manages her stress, \textit{“I'm always trying to walk away from my desk. I need to get something I need to print something, and so I'll walk. I'll get water. I'll go to the bathroom.”}

Other participants decided to take mental breaks instead, shifting their workload from the stressful task at hand to something cognitively less demanding. For instance, P1 reported that taking on a personal task for some time helped him manage his stress: \textit{“I think I went online to check just personal emails since it was the holidays and you know, kind of just like reaching out to people. I think that was like my 10-15 min breather.”} Similarly, while preparing for a foreseeable stressful work task, P6 stated, \textit{“I allow[ed] myself 30 min to relax and slowly set up whatever I need[ed] to set up.”}

\subsubsection{A mixed outlook on sharing stressful experiences}

Five participants also reported that communicating around their stressful experiences helped them better manage their stress. Sharing among colleagues seemed to be the most accessible way to process and let go of these events. For example, P2 noted, \textit{“connecting with other people is really helpful. Even if it's just seeing them and go ask questions, it's helpful.”} Similarly, P6 shared, \textit{“I did share with my coworker how I felt. [...] And he was very reassuring, because he's also experienced it. We both have similar duties in our work titles.”}

However, workplace dynamics can make it difficult for some to openly share their experiences. While P7 had a positive experience communicating with her supervisor, saying, \textit{“I talked to her, I calmed down and she agreed that we're not going to put her on the rotation anymore. My stress was immediately lowered because I felt heard,”} others faced challenges. For instance, P6 and P8 expressed hesitation sharing their feelings with their supervisors. P6 admitted, \textit{“I don't want to. I didn't want to express it to [my supervisor],”} while P8 explained, \textit{“I'm not getting past the whole title.”}

\subsubsection{Environmental factors}

Some of the stress management strategies reported by our participants are not grounded in psychological or organizational research. Instead, they incorporate environmental factors to help alleviate stress. One participant specifically mentioned using sound cues to help cope with his stress. In addition to physical breaks, P4 stated, \textit{“I listen to music while I am working”} while at the office. While working from home, this participant relied on another source of aural stimulation. P4 reported, \textit{“I'll turn on the TV in the background, something that I like, a TV show that I've already watched before.”} Additionally, P2 also reported that she would benefit from changing her visual scenery by working close to a window, as discussed in Section \ref{workplace_design}.

\subsection{Impacts of stress and self-reflections}

Finally, we report on the different impacts of stress in the workplace and worker self-reflections about these events. 

\subsubsection{Emotional and physical responses}

Participants navigated a spectrum of intense and challenging emotions during their experiences, often feeling frustration, hopelessness, and self-doubt. Organizational stress was a prominent theme, with participants expressing frustration over time constraints and misaligned priorities. P6 voiced this clearly, stating, \textit{"It sucks that I have to sacrifice my time or make my schedule work for an event that's not related to me,"} while P3 felt \textit{"frustrated and annoyed"} at the \textit{"lack of respect for my time"} and the unrealistic nature of some requests. Similarly, P5 described a sense of dread, saying, \textit{"I was a little worried because I had already submitted the budget,"} indicating how administrative pressures and management issues amplified her stress. 

Participants also frequently expressed feeling a lack of control when stressed, escalating to feeling incapacitated at times. For instance, some felt a lack of control. P7 stated, \textit{“I guess it was also my feeling of helplessness, looking bad in front of other people, unable to help control the situation.”} This sentiment of disarray was echoed by P2, who explained, \textit{"I am scrambling for answers instead of being in control.”} As a result of their stress, some participants voiced their inability to efficiently manage tasks. For instance, P8 specifically discussed how highly stressful events would make it challenging to go through her normal day of work. She highlighted, \textit{“I can't find the file because I'm stressed out, I'm not able to communicate because I'm stressed out, I'm not able to focus because I'm stressed out,”} which later escalated, \textit{“my heart beats really fast, I start to stutter, I start fidgeting with my hands, my hands get sweaty, and I get into panic mode.”} These accounts highlight how stress can impair both cognitive and physical functioning, compounding the difficulty of navigating already demanding circumstances, and having lasting effects on worker well-being.

Feelings of inadequacy were another significant theme, as participants grappled with self-doubt and a fear of failure. P2 admitted when asked how she felt about her experience, \textit{"and it looks like I don't know what I'm doing. [...] So frustrated and kind of ashamed,"} reflecting a deep insecurity about her performance. P8 echoed this sentiment, describing themselves as feeling \textit{"incompetent"} and that they were \textit{"letting the division down."} These feelings sometimes lead to isolating behaviors to avoid showing vulnerability. P8 stated, \textit{"I stay in my office just because I don't want to be emotional when I go outside"} as she fears students, faculty, and staff members would come to her for her questions because of her role.  

Finally, emotional exhaustion arose as a recurring experience. P7 felt \textit{"emotionally drained"} and noted, \textit{"I guess I was more stressed because she [her supervisor] has done this numerous times,"} highlighting the toll of repeated stressors and the place of organizational stress in the workplace. P4 described a sense of hopelessness, saying, \textit{"I felt so hopeless,"} which was compounded by his deep-seated fear of disappointing others: \textit{"I hate letting people down."} Together, these themes underscore the heavy emotional burden participants carried due to persistent stress from external demands and internal pressures in the workplace.

\subsubsection{Introspection and self-reflections}

Despite strong emotional reactions, these experiences allowed participants to further reflect on their stress at work. Several participants expressed a positive outlook when revisiting these moments, even if they were difficult at the time. For instance, P4 shared, \textit{“while that was rather negative, looking back at it, it's a positive experience.”} Similarly, P6 highlighted a consistent effort to frame challenges positively, \textit{“I like to take things as a positive experience, usually because I'm always learning something about myself, even if it's a bad time.”} These reflections underscore significant personal growth and reveal that stress, while challenging, can serve as a powerful driver for self-development.

Participation in the data collection process prior to this study appeared to catalyze self-awareness and introspection, though effects on daily life varied. For instance, P1 found the data collection impactful but questioned its long-term influence. He specifically highlighted that interaction with his data in the moment was helpful, but these self-reflections do not exist anymore, stating, \textit{"I got a chance to do feedback and all those things which were really helpful [at the time] [...] But since it was so far removed in a way that I don't really feel it anymore. The recollection of all that I did during the study [n.b., longitudinal data collection] was a daily, relevant thing."} Meanwhile other participants reported clearer benefits still holding true long after the data collection process was finished. For example, P4 found that his participation made him \textit{“more mindful of my stress,”} while P6 highlighted long-term benefits, stating, \textit{“I'm more self aware of what my triggers could be for stress.”}

Increased self-awareness often translated into actionable changes. For instance, P8 shared a transformative realization, stating, \textit{"some of the questions that were asked [during the data collection process] really made me think of how I view myself as a manager […] before I would bottle it up [...] I took it home and would yell at my kids, my husband, at the cat, at the dog, get stressed out."} She attributed the longitudinal data collection’s introspective EMAs to helping her reflect on stressful experiences and manage her stress more effectively. Similarly, P2 discussed deeper insights into the source of her stress while participating in the data collection phase, allowing her to reflect on how her experience stress: \textit{“some introspection on what's really causing the stress in this event.”} While some participants found accessing and reflecting on their stress data helpful, others noted that being stressed distracted them and made it challenging from engaging with their data effectively. For example, preoccupied by completing a stressful task at hand, P4 explained that he couldn’t think to reflect on his stress data: \textit{“I wasn’t looking at my Fitbit to see what my heart rate was at the time.”}

\section{Discussion}

Workplace stress is a growing concern, with technological advances to support stress management. However, much of this research overlooks the unique needs and experiences of office workers, whose stress is shaped by a variety of complex factors. This study aims to explore these experiences, providing insights into designing more user-centered solutions that effectively address the varied needs of office workers, enabling better workplace stress management.

\paragraph{Personalizing Interventions}
Our participants’ stress experiences varied notably depending on psychological job demands and demographic factors such as gender and age. Higher-psychological-demand roles often reported heightened organizational stress tied to unrealistic deadlines and administrative hurdles. Conversely, those with lower psychological demands indicated more personal-life spillover, suggesting different expectations in one’s job role does not necessarily reduce overall stress but may shift its locus. To some extent, we also note differences in stress experiences across genders. For instance, two female participants specifically mentioned how inadequate workspace design increased their stress.
Our findings suggest stress management interventions may better adapt to specific demand profiles, as suggested by prior research \cite{richardson_effects_2008}. For instance, workers with high psychological-demand roles (particularly administrative workers) might benefit from interventions tied to their work commitments and space, whereas workers with lower psychological-demand roles could benefit from on-demand interventions that can be used outside the workspace. Similarly, female workers might benefit more from interventions tied to environmental-related stressors than male workers. While our limited sample cannot yield generalized conclusions, our findings suggest that stress management interventions can be tailored to different worker job roles, work demands, and demographics. We further discuss implications for designing stress management technologies in the following sections.

\paragraph{Social Barriers, Hierarchies, and Trust in Stress Disclosure}
Participants’ experiences highlight the importance of social connections in mitigating workplace stress, aligning with prior evidence that informal interactions among colleagues can enhance well-being \cite{begemann_enabling_2024}. Yet, our findings also underscore structural barriers that constrain open disclosure. For instance, hierarchical relationships can discourage employees (particularly administrative workers) from acknowledging vulnerabilities. In some cases, even the physical openness of shared offices can increase apprehension about visibly engaging with stress-management tools. These observations mirror concerns about workplace “surveillance” practices, wherein employees worry that data on their stress levels could be misread or misused, ultimately reducing trust rather than fostering well-being \cite{glavin_private_2024, kawakami_sensing_2023}. Making individual metrics publicly visible can induce self-censorship, as workers focus on avoiding negative perceptions rather than seeking genuine support \cite{kawakami_sensing_2023, das_swain_sensible_2024}. These challenges highlight the tension between helpful transparency and fear of stigma. Rather than simply calling for better privacy safeguards, our insights illustrate specific design levers for mitigating the risks of open disclosure. We observed that different job titles and supervisory relationships either facilitated or impeded stress discussions. This points to a design opportunity for discreet or anonymous approaches that promote help-seeking without risking repercussions. For example, employees with more demanding roles and those with higher privacy-needs could benefit from conversational agents with data anonymization to share stress-related challenges. 

\paragraph{Expanding Stress Management Interventions to the Physical Environment}
While organizational factors remain a prominent source of workplace stress \cite{bolliger_sources_2022}, workplace design and operations surfaced as a compelling avenue for mitigating these stressors. Workers’ accounts illustrate that workplace environmental demands are prevalent \cite{awada_ten_2023}. Poor IEQ factors such as noise \cite{Toftum_2012, sander_open-plan_2021, Choi_2015}, lighting \cite{Thach_2020}, and indoor air quality \cite{Thach_2020, Choi_2015}, have been shown to contribute significantly to worker stress. Scholars highlight that IEQ interventions can support worker stress management, particularly when supported by multimodal sensing (e.g., capturing environmental data in tandem with wearable or usage metrics) \cite{zhao_mediated_2017, zhao_mediated_2022}. Yet, the lived experiences of office workers exposed to these interventions remain underexplored, particularly concerning how they interact with interpersonal dynamics and work processes. Notably, some participants actively leverage their existing workplace environments to manage stress, hinting at the opportunity for adaptive, context-sensitive solutions. Recent work in robotics further underscores the efficacy of adaptive workspaces for tackling organizational (e.g., disruptions) and environmental (e.g., acoustic) constraints \cite{nguyen_adaptive_2024, nguyen_exploring_2021}. Future research could therefore investigate not only the reactive nature of environmental interventions (e.g., changing environmental conditions when users are stressed) but also preventive measures (e.g., changing environmental conditions and creating cues to encourage breaks before stress accumulates). In doing so, it is crucial to assess how various dimensions (e.g., workplace layout (open-plan vs. closed-office), job role demands, and organizational culture, intervention modalities, tangibility) mediate the effectiveness of these interventions.

\paragraph{Worker Context and Cognitive Load}
Another important avenue in stress management interventions is the interplay between stress and cognitive load. Many of our participants described feeling paralyzed by stress, largely due to distractions that align with what cognitive load theory describes as \textit{extraneous load} (e.g., interruptions) \cite{sweller_chapter_2011, norman_design_2013}. These observations highlight how stressors - particularly those that demand immediate attention - can conflict with the modality of an intervention (e.g., visual prompts or auditory alerts), thereby reducing its effectiveness or even exacerbating stress \cite{norman_design_2013}. In addition to increasing cognitive load, stress can persist over time, making workers less responsive to interventions at best, or compound the negative effects of stress at worst. Although some research has explored the efficacy of timing interventions \cite{suh_toward_2024, howe_design_2022, clarke_mstress_2017, lustrek_designing_2023}, our findings suggest that stress remains disabling long after the precipitating event. Future work should investigate how to optimize stress management intervention timing and modalities to minimize extraneous cognitive load and avoid contributing additional, unnecessary stress.

\paragraph{Designing for Self-Reflection and Long-Term Engagement}
Finally, our interviews brought to light the importance of self-reflection in worker stress management. Despite the passive data collection nature of our study, participants specifically discussed how regularly reporting their mental and physical health data positively affected them. These findings are consistent with current research in personal informatics. While reflecting on personal data can be challenging \cite{Rapp_2016, Bentvelzen_2021}, recent research highlights the benefits of self-reflection on worker well-being \cite{jorke_pearl_2023} and health \cite{epstein_mapping_2020}. However, this study also revealed that the long-term effects of reflecting on past personal data were mixed. Future research should therefore explore how self-reflection technologies and sustained engagement with personal data influence the effectiveness of workplace stress management interventions, over time.

All in all, we highlight key takeaways to support future user-centered stress management systems: 

\begin{enumerate}
    \item \textbf{Stress is complex and heavily worker-dependent.} Office workers face a variety of stressors from organizational, interpersonal, and individual sources. Workers with high job demands might benefit from interventions seamlessly integrated into their daily workflow, while those with lower demands may prefer flexible, on-demand solutions that can be used outside the workspace.
    \item \textbf{Stress management interventions cannot be bound to digital interfaces alone.} The adaptive workplace can be leveraged for stress management, providing an additional layer of physical and intangible interventions. For example, changes in workplace design (i.e., space, partitions) and operations (i.e., noise, lighting) can be used to help workers manage their stress.
    \item \textbf{Timing and modalities of interventions matter.} Stress can linger long after a triggering event, and stress management interventions can introduce additional extraneous cognitive load. Poorly timed or overly intrusive interventions may worsen stress.
    \item \textbf{Stress is not always negative.} Designing interventions that reduce distress while promoting positive stress (i.e., eustress) can improve worker well-being.
    \item \textbf{Improve data awareness and self-reflection.} Reflecting on personal data is an effective way to manage and reduce stress. Interventions should be designed to promote and encourage sustained self-reflection, over time. 
\end{enumerate}

Our study presents several limitations. First, the small sample size limits the generality of our findings as the perspectives captured may not represent broader workplace trends. Second, our participants were all affiliated with USC, which may omit stressors and coping mechanisms unique to workers from other industries or professional contexts. Factors such as organizational culture, job roles, and industry-specific dynamics likely influence how stress manifests and is managed. We partially address this limitation by having recruited workers with different work and stress demands, as well as varied demographics (Table \ref{tab:demographics}). Additionally, our study relied on retrospective accounts where participants reflected on past stressful experiences. While this approach provided valuable insights, richer findings could emerge by exploring how office workers reflect and engage with their actual data.

\section{Conclusion}

In this study, we contextualized stress in the workplace leveraging insights from semi-structured interviews. This work sheds light on the multifaceted nature of workplace stress through diverse office worker experiences. Our findings highlight the complexity and interplay of workplace stressors, and how they shape worker stress. We emphasize the need to leverage environmental and contextual factors within the workplace itself for stress management interventions, the critical role of timing and modality in implementing these interventions, and the benefits of self-reflection on personal data. Additionally, we highlight the dual nature of stress—not only as a challenge but also as a catalyst for growth. Researchers can use these insights to design more user-centered and effective stress management interventions.

\begin{acks}
This publication was supported by the National Science Foundation under Grant No. 2204942. The authors are solely responsible for the content, which does not reflect the official views of the National Science Foundation. This study was conducted according to the guidelines of the Declaration of Helsinki and approved by the Institutional Review Board of the University of Southern California (UP-22-00548, Effective Approval Date: 13 July 2022). 
Additionally, Dr. Gale Lucas has been supported by the U.S. Army Research Office under Grant No. W911NF2020053. Any findings, opinions, conclusions, or recommendations presented in this study are those of the authors and do not reflect the views of the National Science Foundation or the U.S. Army Research Office necessarily.
\end{acks}

\bibliographystyle{ACM-Reference-Format}
\bibliography{CHI_bib}

\appendix

\section{Participant summary}

Participant demographics and psychosocial job characteristics are included in Table \ref{tab:demographics}. The psychosocial job characteristics scale, based on the JCQ \cite{karasek_job_1998}, measures "Decision authority" and "Psychological demands" on a normalized scale from -18 to 18. A lower score indicates minimal decision authority or psychological demands, while a higher score reflects greater presence of these aspects within a job role. For example, higher decision authority and psychological demands are indicative of an active job. On the other hand, high psychological demands and low decision authority are indicative of high strain.

\begin{table*}[htb]
  \caption{Participant summary. We included the participants demographics as well as their psychosocial job characteristics. All participants were associated with USC.}
  \label{tab:demographics}
  \resizebox{\textwidth}{!}{
  \begin{tabular}{c c c c c c c c}
    \toprule
    ID & Age & Race & \makecell{Hispanic \\ or Latino} & Gender & Work Type & \makecell{Decision \\ authority} & \makecell{Psychological \\ demands} \\
    \midrule
    P1 & 46 & Native Hawaiian or other Pacific Islander & Yes & M & Administrative Assistant & -2 & 6 \\
    P2 & 42 & Black or African American & No & F & Assistant Professor & 2 & -1 \\
    P3 & 45 & White & No & M & Graduate Programs Manager & 10 & 8 \\
    P4 & 41 & Asian & No & M & Student Services Advisor & 6 & 1 \\
    P5 & 56 & White & No & F & Senior Research Administrator & -6 & 1 \\
    P6 & 26 & White & Yes & F, NB & Administrative Assistant II & 10 & 2 \\
    P7 & 52 & White & No & F & Senior Administrator & 2 & 4 \\
    P8 & 47 & White & Yes & F & Department Business Manager & 2 & 4 \\
    \bottomrule
  \end{tabular}}
\end{table*}

\section{Semi-structured interview guide} \label{itw_guide}
The following section outlines the interview guide used for this project. We used initial questions to guide the conversion, and probe questions were asked to elicit further insights into the participants' experience.

\subsection{Initial questions}

Recall a couple of times during the months in which the study occurred that you felt particularly stressed while you were at your workstation. You may think of more than two examples.
\begin{enumerate}
    \item Can you tell us in as much detail as you can what led you to feeling stressed?
    \item During these stressful events, what actions did you take to remediate your stress, if any?
    \item Right after these stressful events happened, did you ever share about how you felt with others?
\end{enumerate}

\subsection{Probe questions}

\begin{enumerate}
    \item Where did this event occur?
    \item Did this specific event also happen while you were working somewhere else?
    \item When did this event happen? [ask participants to give a date and time, if possible]
    \item How often would the event you just described happen? 
    \item Did this event happen regularly or at a specific instance in time?
    \item What kind of activity were you doing during this stressful event? Was it work-related or non-work related?
    \item Do you believe that the activity you were participating in impacted your stress levels? If so, how?
    \item Do you believe that your environment had an impact on your stress? (lighting, window view, temperature, etc. if question is unclear)
    \item Do you believe that factors unrelated to your environment had an impact on your stress? (meetings, family issues, etc. if question is unclear)
    \item How helpful were the steps you took to remediate/manage your stress? (if unclear: Did you manage to remediate your stress? Is there anything that worked well? Anything that did not work?) 
    \item Looking back at this event now, do you think you could have managed your stress differently? If so, how?
    \item How did you feel emotionally during this event? 
    \item How exactly did you know you were stressed? Are there any factors/indicators that led you to feel you were experiencing stress?
    \item Did participating in the study impact how this stressful event unfolded? If so, how?
    \item Was there anything different about this event than usual? 
    \item What did you learn about yourself and your stress after this stressful event?
    \item Was this event a positive or negative experience? 
    \item What kind of information did you share?
    \item Who did you share this information with?
    \item Why did you decide to share this information? Did you believe it helped you in any way?
    \item Would you want other people in your work circles to know when you felt stressed? Would you like them to understand why?
    \item Would you want to know when other people in your work circles felt stressed? Would you like to know why?
\end{enumerate}

\balance

\end{document}